\documentclass[12pt]{article}
 \usepackage[cp1251]{inputenc} %- для WiN
 \usepackage[english]{babel}
 \usepackage{amsmath, latexsym, amssymb, bm, array, graphics, eucal,
 amsfonts,}%mathscr
 \makeatletter
 \renewcommand{\@biblabel}[1]{#1.\hfill}

\renewcommand{\div}{\mathop{\rm div}}
\newcommand{\mc}[1]{\mathcal{#1}}
\newcommand{\E}{\mc{E}}
 \makeatother
 
\usepackage[dvips]{graphicx}

\begin{document}
 \thispagestyle{empty}
 \renewcommand{\abstractname}{\ }
 \large
 \renewcommand{\refname}{\begin{center} REFERENCES\end{center}}
\newcommand{\const}{\mathop{\rm const\, }}
 \begin{center}
\bf LANDAU DIAMAGNETISM OF DEGENERATE COLLISIONAL PLASMA
\end{center}\medskip
\begin{center}
  \bf  A. V. Latyshev\footnote{$avlatyshev@mail.ru$} and
  A. A. Yushkanov\footnote{$yushkanov@inbox.ru$}
\end{center}\medskip

\begin{center}
{\it Faculty of Physics and Mathematics,\\ Moscow State Regional
University,  105005,\\ Moscow, Radio st., 10--A}
\end{center}\medskip

\begin{abstract}
For the first time the kinetic description of  Landau diamagnetism
for degenerate collisional plasma is given.
The correct expression for transverse electric conductivity of the quantum
plasma, found by authors (see arxiv:1002.1017 [math-ph] 4 Feb
2010) is used. In work S. Dattagupta, A.M. Jayannavar and N.
Kumar [Current science, V. 80, No. 7, 10 April, 2001]
was discussed the important problem of  dissipation (collisions)
influence on  Landau diamagnetism. The analysis of this problem
is given with
the use of exact expression for transverse conductivity of quantum plasma.

\medskip

\medskip

electron gas

\end{abstract}

\begin{center}\bf
  1. Introduction
\end{center}

Magnetisation of electron gas in a weak magnetic fields
compounds of two independent parts (see, for example, \cite {Landau5}):
from the paramagnetic magne\-ti\-sa\-tion connected
with own (spin) magnetic
momentum of electrons ({\it Pauli's para\-mag\-netism}, W. Pauli, 1927)
and from the diamagnetic mag\-ne\-ti\-sation connected with
quantization of
orbital movement of electrons in a magnetic field ({\it
Landau diamagnetism}, L. D. Landau, 1930).

Landau diamagnetism  was considered till now for a gas of the free
elect\-rons. It has been thus shown, that together with original
approach develo\-ped by Landau, expression for diamagnetism of electron
gas can be obtained on the basis of the kinetic approach
\cite {Silin}.

The kinetic method gives opportunity to calculate the trans\-verse
die\-lect\-ric permeability.  On the basis of this quantity its possible
to obtain
the value of the diamagnetic response.

However such calculations till now
were carried out only for collisionalless case. The matter is that
correct expression for the transverse dielectric
permeability of quantum plasma existed till
 now only for gas of the free
electrons. Expression known till now for the transverse dielectric
perme\-abi\-lity in  a collisional case gave incorrect transition to
the classical case \cite {Kliewer}. So this expression  were accordingly
incorrect.

In work \cite {LY}  for the first time the expression   for
the quantum transverse dielectric
permeability of collisional degenerate plasma has been derived. The
obtained in \cite {LY} expression for
trans\-ver\-se dielectric permeability satisfies
to the necessary requirements of com\-pa\-ti\-bility.

Central result from \cite{Datta} connects the mean orbital
magnetic moment, a thermodynamic property, with the electrical
resistivity, which characterizes transport properties of
material. In this work
was discussed the important problem of  dissipation (collisions)
influence on  Landau diamagnetism. The analysis of this problem
is given with
use of exact expression of transverse conductivity of quantum plasma.

In work \cite{Kumar} is shown that a classical system of charged
particles moving on a finite but unbounded surface (of a sphere)
has a nonzero orbital diamagnetic moment which can be large.
Here is considered a non-degenerate system with the degeneracy
temperarure much smaller than the room temperature, as in the
case of a doped high-mobility semiconductor.

In the present work with use of correct expression for the
transverse conductivity
\cite {LY} the kinetic description  of Landau diamagnetism
for degene\-ra\-te collisional plasma
for the first time is given.

\begin{center}
  \bf 2. The general expression for a magnetic susceptibility
\end{center}

Magnetization vector $\mathbf{M}$ of electron plasma
is connected with current density $\mathbf{j}$ by the following
expression \cite {Landau8}
$$
{\bf j}=c\, {\rm rot}\,\mathbf{M},
$$
where $c$ is the light velocity.

Magnetization vector $\mathbf{M}$ and a magnetic field
strength
$\mathbf{H}=\rm rot \mathbf{A}$ are connected by the expression
$$
{\bf M}=\chi\,{\bf H}=\chi\,{\rm rot}\,{\bf A},
$$
where $\chi$ is the magnetic susceptibility, $\mathbf{A}$ is the
vector potential.

From these two equalities for current density we have
$$
\mathbf{j}={c}\, {\rm rot}\,\mathbf{M}=
c\,\chi {\rm rot}\, \big({\rm rot}\,\mathbf{A}\big)=
c\,\chi\, \big[{\bf \nabla}({\bf \nabla}{\bf A})-
{\mathbf{\triangle}}\mathbf{A}\big].
$$

Here $\Delta$ is the Laplace operator.

Let the scalar potential is equal to zero.
Vector potential we take ortho\-gonal
to the direction of a wave vector $\mathbf{k}$
($\mathbf{k}\mathbf{A}=0$) in the form of a  harmonic wave
$$
\mathbf{A}(\mathbf{r},t)=\mathbf{A}_0
e^{i(\mathbf{k} \mathbf{r}-\omega t)}.
$$

Such vector field is solenoidal
$$
\div \mathbf{A}=\nabla\mathbf{A}=0.
$$

Hence, for current density we receive equality
$$
{\bf j}=-c\,\chi \Delta \mathbf{A}=c\,\chi\,k^2 \mathbf{A}.
\eqno{(1.1)}
$$

On the other hand, connection of electric field  $\mathbf{E}$ and vector
potential $\mathbf{A}$
$$
\mathbf{E}=-\dfrac{1}{c}\dfrac{\partial \mathbf{A}}{\partial t}=
\dfrac{i\omega}{c}\mathbf{A}
$$
leads to the relation
$$
\mathbf{j}=\sigma_{tr}\mathbf{E}=\sigma_{tr}\dfrac{i\omega}{c}
\mathbf{A},
\eqno{(1.2)}
$$
where $\sigma_{tr}$ is the transverse electric conductivity.

For our case from (1.1) and (1.2) we obtain
following expression for the magnetic susceptibility
$$
\chi=\dfrac{i\omega}{c^2 k^2}\sigma_{tr}.
\eqno{(1.3)}
$$

\begin{center}
  \bf 3. Magnetic susceptibility of degenerate plasma
\end{center}

Landau diamagnetism  in collisionless  plasma is usually
 defined as a magnetic susceptibility in a static limit
for a homogeneous external magnetic field.
Thus the diamagnetism value can be found by means of (1.1)
through two  non-commutative limits
$$
\chi_L=\lim\limits_{k\to 0}\Big[\lim\limits_{\omega\to 0}^{}
\chi(\omega,k,\nu=0)\Big].
\eqno{(2.1)}
$$

Here $\nu$ is the effective electron collision frequency.

In collisionless plasma %$ \omega=0$, $k\to 0$
this expression (2.1) should lead to the known formula for Landau
diamagnetism
$$
\chi_L=-\dfrac{1}{3}
\left(\dfrac{e\hbar}{2m c}\right)^2\dfrac{p_Fm}{\pi^2\hbar^3}.
\eqno{(2.2)}
$$

Here $e$ and $m$ are the electron charge and mass  accordingly,
$\hbar$ is the Planck's constant, $p_F $ is the electron momentum on
Fermi's surfaces which is considered as spherical.

We take the Cartesian coordinates system with an axis $x$
directed lengthways vector $\bf k$, and an axis $y$ along the
vector $\bf A$.
Then expression for transverse conductivity of degenerate collisional
plasma is defined by the general formula \cite {LY}
$$
\sigma_{tr}=\dfrac{e^22m^3}{\omega (2\pi\hbar)^3}\int
\dfrac{(f_F^+-f_F^-)/\hbar+(\omega-kv_x)\delta(\E_F-\E)}
{\nu-i\omega+ikv_x}v_y^2d^3v,
\eqno{(2.3)}
$$
where $\E_F$ is the kinetic energy of electrons on Fermi's surface,
$\delta(x)$ is the Dirac's delta function,
$$
\E_F=\dfrac{p_F^2}{2m}=\dfrac{mv_F^2}{2},\qquad
f_F^{\pm}\equiv \Theta(\E_F-\E^{\pm})\equiv \Theta^{\pm},
$$
$$
\E^{\pm}=\E^{\pm}(k)=
\dfrac{\Big(p_x\mp\dfrac{k\hbar}{2}\Big)^2}{2m}+\dfrac{p_y^2+p_z^2}{2m}=
$$

$$
\hspace{5cm}
=\dfrac{m}{2}\Big[\Big(v_x\mp\dfrac{\hbar
k}{2m}\Big)^2+v_y^2+v_z^2\Big],
$$
$\Theta(x)$ is the Heaviside function,
$$
\Theta(x)=\left\{\begin{array}{c}
                   1,\qquad x>0, \\
                   0, \qquad x<0.
                 \end{array}
\right.
$$

We will present the formula (2.3) in the form of the sum of classical
and quantum components
$$
\sigma_{tr}=\sigma_{tr}^{\rm classic}+\sigma_{tr}^{\rm quant},
\eqno{(2.4)}
$$
where
$$
\sigma_{tr}^{\rm classic}=\dfrac{e^22m^3}{(2\pi\hbar)^3}\int
\dfrac{\delta(\E_0-\E)v_y^2}{\nu-i\omega+ikv_x}d^3v
\eqno{(2.5)}
$$
and
$$
\sigma_{tr}^{\rm quant}=\dfrac{e^22m^3}{\omega(2\pi\hbar)^3}\int
\dfrac{-kv_x+(\Theta^+-\Theta^-)/\hbar}{\nu-i\omega+ikv_x}v_y^2d^3v.
\eqno{(2.6)}
$$

On the basis of equalities (2.4) -- (2.6) we  write down the corresponding
equalities for a magnetic susceptibility
$$
\chi=\chi^{\rm classic}+\chi^{\rm quant},
\eqno{(2.7)}
$$
where
$$
\chi^{\rm classic}=\dfrac{2i\omega e^2m^3}{c^2k^2(2\pi\hbar)^3}
\int\dfrac{\delta(\E_0-\E)v_y^2}{\nu-i\omega+ikv_x}d^3v
\eqno{(2.8)}
$$
and
$$
\chi^{\rm quant}=\dfrac{2ie^2m^3}{c^2k^2(2\pi\hbar)^3}
\int
\dfrac{-kv_x+(\Theta^+-\Theta^-)/\hbar}{\nu-i\omega+ikv_x}v_y^2d^3v.
\eqno{(2.9)}
$$

\begin{center}
\bf 3. Transverse conductivity analysis. % decomposition on degrees of wave number.
Diamagnetic properties of metal
\end{center}

Let's decompose expression (2.6) by degrees of wave number $k$.

 The functions $\Theta^{\pm}$ may be presented in the form
$$
f_F^{\pm}(k)=\Theta^{\pm}(k)=
\Theta(\E_F-\E-\dfrac{\hbar^2k^2}{8m}\pm\dfrac{v_x\hbar}{2}k).
$$

It is clear, that
$$
f_F^{\pm}(0)=\Theta^{\pm}(0)=\Theta(\E_F-\E).
$$

The first derivative of Fermi --- Dirac distribution function is equal to
$$
\dfrac{\partial \Theta^{\pm}(k)}{\partial k}=
\delta\Big(\E_F-\E-\dfrac{\hbar^2k^2}{8m}\pm\dfrac{v_x\hbar}{2}k\Big)
\Big(-\dfrac{\hbar^2}{4m}k\pm\dfrac{v_x\hbar}{2}\Big).
$$

From here follows
$$
\dfrac{\partial \Theta^{\pm}(0)}{\partial k}=
\delta\Big(\E_F-\E\Big)
\Big(\pm\dfrac{v_x\hbar}{2}\Big),
$$
where $\delta(x)$ is the Dirac's delta function.

The second derivative of Fermi --- Dirac distribution function is
equal to
$$
\dfrac{\partial^2 \Theta^{\pm}(k)}{\partial k^2}=
\delta'\Big(\E_F-\E-\dfrac{\hbar^2k^2}{8m}\pm\dfrac{v_x\hbar}{2}k\Big)
\Big(-\dfrac{\hbar^2}{4m}k\pm\dfrac{v_x\hbar}{2}\Big)^2+
$$
$$
+\delta\Big(\E_F-\E-\dfrac{\hbar^2k^2}{8m}\pm\dfrac{v_x\hbar}{2}k\Big)
\Big(-\dfrac{\hbar^2}{4m}\Big).
$$

From here we find its value, when $k=0$
$$
\dfrac{\partial^2 \Theta^{\pm}(0)}{\partial k^2}=
\delta'\Big(\E_F-\E\Big)\dfrac{v_x^2\hbar^2}{4}-\dfrac{\hbar^2}{4m}
\delta\Big(\E_F-\E\Big).
$$

The third derivative of Fermi --- Dirac distribution function is
equal to
$$
\dfrac{\partial^3 \Theta^{\pm}(k)}{\partial k^3}=
\delta''\Big(\E_F-\E-\dfrac{\hbar^2k^2}{8m}\pm\dfrac{v_x\hbar}{2}k\Big)
\Big(-\dfrac{\hbar^2}{4m}k\pm\dfrac{v_x\hbar}{2}\Big)^3+
$$
$$
+3\delta'\Big(\E_F-\E-\dfrac{\hbar^2k^2}{8m}\pm\dfrac{v_x\hbar}{2}k\Big)
\Big(-\dfrac{\hbar^2}{4m}k\pm\dfrac{v_x\hbar}{2}\Big)
\Big(-\dfrac{\hbar^2}{4m}\Big).
$$

From here we find
$$
\dfrac{\partial^3 \Theta^{\pm}(0)}{\partial k^3}=\pm\dfrac{v_x^3\hbar^3}{8}
\delta''\Big(\E_F-\E\Big)\mp
\dfrac{3v_x\hbar^3}{8m}\delta'\Big(\E_F-\E\Big).
$$

By means of the found derivatives we receive decomposition of the
Fermi --- Dirac  distribution function
$$
\Theta^{\pm}(k)=\Theta(\E_F-\E)\pm
\dfrac{v_x\hbar}{2}\delta(\E_F-\E)k+\hspace{3cm}
$$
$$\hspace{1cm}
+\Big[\dfrac{v_x^2\hbar^2}{4}\delta'(\E_F-\E)-\dfrac{\hbar^2}{4m}
\delta(\E_F-\E)\Big]\dfrac{k^2}{2}\pm
$$
$$\hspace{2cm}
\pm\Big[\dfrac{v_x^3\hbar^3}{8}\delta''(\E_F-\E)-
\dfrac{3v_x\hbar^3}{8m}\delta'(\E_F-\E)\Big]\dfrac{k^3}{6}.
$$

The difference of these decompositions,
devided  by Planck's constant, is equal to
$$
\dfrac{\Theta^+(k)-\Theta^-(k)}{\hbar}=v_x\delta(\E_F-\E)k+$$$$+
\Big[v_x^3\hbar^2\delta''(\E_F-\E)-3\dfrac{v_x\hbar^2}{m}
\delta'(\E_F-\E)\Big]\dfrac{k^3}{24}.
$$

The second term of numerator from integrand (2.6) is equal to
$$
\dfrac{\Theta^+(k)-\Theta^-(k)}{\hbar}-kv_x\delta(\mc{E}_0-\mc{E})
=$$$$=\dfrac{\hbar^2k^3}{24}
\Big[v_x^3\delta''(\mc{E}_0-\mc{E})-3\dfrac{v_x}{m}
\delta'(\mc{E}_0-\mc{E})\Big].
$$

Now for quantum conductivity (2.6) we have the following expression
$$
\sigma_{tr}^{\rm quant}=\dfrac{e^2m^3\hbar^2k^3}{12\omega(2\pi\hbar)^3}
\int\dfrac{v_x^2\delta''(\E_F-\E)-({3}/{m})
\delta'(\E_F-\E)}{\nu-i\omega+ikv_x}v_xv_y^2\;d^3v.
\eqno{(3.1)}
$$

According to (2.8) we will write expression for the classical diamagnetic
susceptibilities
$$
\chi^{\rm classic}=i\dfrac{2e^2m^3\omega}{(2\pi\hbar)^3 c^2 k^2}\int
\dfrac{\delta(\E_0-\E)v_y^2}{\nu-i \omega +ikv_x}d^3v,
\eqno{(3.2)}
$$
and according to (2.9) we will write expression for the quantum diamagnetic
susceptibility
$$
\chi^{\rm quant}=i\dfrac{e^2m^3\hbar^2k}{12(2\pi\hbar)^3c^2}\int
\dfrac{v_x^2\delta''(\mc{E}_0-\mc{E})-({3}/{m})
\delta'(\mc{E}_0-\mc{E})}{\nu-i\omega+ikv_x}v_xv_y^2\;d^3v.
\eqno{(3.3)}
$$

From the expression (3.2) we see, that
$$
\lim\limits_{\omega\to 0}^{}\chi^{\rm classic}=0.
\eqno{(3.4)}
$$

The relation (3.4) means, that magnetisation of the classical
gas is equal to zero.
%\medskip

According to (1.3) and taking into account deduced earlier
(see \cite {LY}) formulas for
transverse conductivity of quantum degenerate plasma we receive
the following expression for the magnetic susceptibility
$$
\dfrac{\chi}{\chi_L}=-\dfrac{3x}{q^2}\int\limits_{-1}^{1}
\dfrac{(1-t^2)dt}{qt-z}+\dfrac{3}{q}\int\limits_{-1}^{1}
\dfrac{t(1-t^2)dt}{qt-z}+$$$$+
\dfrac{3}{4}
\int\limits_{-1}^{1}\dfrac{(1-t^2)^2dt}{(qt-z)^2-q^4/4},
\eqno{(3.5)}
$$
where $ \chi_L $ is the value of Landau diamagnetism entered
by equality (2.2).

Here for convenience dimensionless variables are entered
$$
z=\dfrac{\omega+i \nu}{k_Fv_F}=x+iy, \quad
x=\dfrac{\omega}{k_Fv_F}, \quad y=\dfrac{\nu}{k_Fv_F}, \quad
q=\dfrac{k}{k_F},
$$
where $k_F$ is the Fermi's wave number,
$k_F=\dfrac{mv_F}{\hbar}=\dfrac{p_F}{\hbar}$, $p_F$ is the
electron momentum on Fermi's surface.

The formula (3.5) will be used  for the graphic
investigation of a magnetic susceptibility.

\begin{center}
\bf 4. Landau diamagnetic susceptibility
\end{center}

Landau's diamagnetic susceptibility of collisionless
plasma  we will define the following double non-commutative limit
$$
\chi_{L}(\nu=0)=\lim\limits_{k\to 0}
\Big[\lim\limits_{\omega\to 0}\chi(\omega,k,\nu=0)\Big].
\eqno{(4.1)}
$$

Let's show, that with the use of the relation (3.3)
the equality (4.1) leads to known Landau's formula (2.2).

Now we will consider collisionless plasma, i.e. we will put
$\nu=0$. Besides, we will consider a static limit, having put $\omega=0$.
Expression (3.3) becomes simpler thus:
$$
\chi^{\rm quant}=\dfrac{e^2m^3}{96\pi^3\hbar c^2}\int
\Big[v_x^2\delta''(\E_F-\E)-
\frac{3}{m}\delta'(\E_F-\E)\Big]
v_y^2\;d^3v.
\eqno{(4.2)}
$$

Calculating integral from (4.2) in spherical system
of coordinates, we obtain
$$
J=\int\Big[v_x^2\delta''(\E_F-\E)-
\frac{3}{m}\delta'(\E_F-\E)\Big]
v_y^2\;d^3v=J_1-\dfrac{3}{m}J_2,
$$
where
$$
J_1=\dfrac{4\pi}{15}
\int\limits_{0}^{\infty}v^6\delta''(\E_F-\E)\;dv,
$$
$$
J_2=\dfrac{4\pi}{3}\int\limits_{0}^{\infty}v^4\delta'(\E_F-\E)\;dv.
$$

%We will calculate integrals $J_1$ and $J_2$  by data to integration on
%energy.
 After replacement of a variable of integration
$v=\sqrt{2\E/m}$, we have
$$
J_1=\dfrac{16\sqrt{2}\pi}{15m^{3/2}}\int\limits_{0}^{\infty}
\delta''(\E_F-\E)\E^{5/2}\;d\E=\dfrac{4\pi v_F}{m^3},
$$
$$
J_2=\dfrac{8\sqrt{2}\pi}{3m^{3/2}}\int\limits_{0}^{\infty}
\delta'(\E_F-\E)\mc{E}^{3/2}\;d\E=\dfrac{4\pi v_F}{m^2}.
$$

Considering that
$$
J_1-\dfrac{3}{m}J_2=-\dfrac{8\pi v_F}{m^3},
$$

we obtain expression for a magnetic susceptibility
$$
\chi_L=-\dfrac{e^2v_F}{12\pi^2\hbar c^2}.
\eqno{(4.3)}
$$

It is easy to check up, that expression (4.3) exactly
coincides with the known Landau  expression (2.2).

\begin{center}
  \bf 5. Diamagnetic susceptibility in collisional plasma
\end{center}

Let's consider the Landau magnetic susceptibility in collisional
plasma, i.e. at $\nu\ne 0$. From the formula (3.3) it follows, that
$$
\chi_L^{\rm quant}(\omega,\nu\ne 0)=\lim\limits_{k\to 0}
\chi^{\rm quant}(\omega,\nu,k)=0.
\eqno{(5.1)}
$$

Let's consider the diamagnetic susceptibility $\chi$ at finite
values of
wave vector $\bf k$.

\begin{figure}[ht]
\begin{flushleft}
\includegraphics[width=17.0cm, height=12cm]{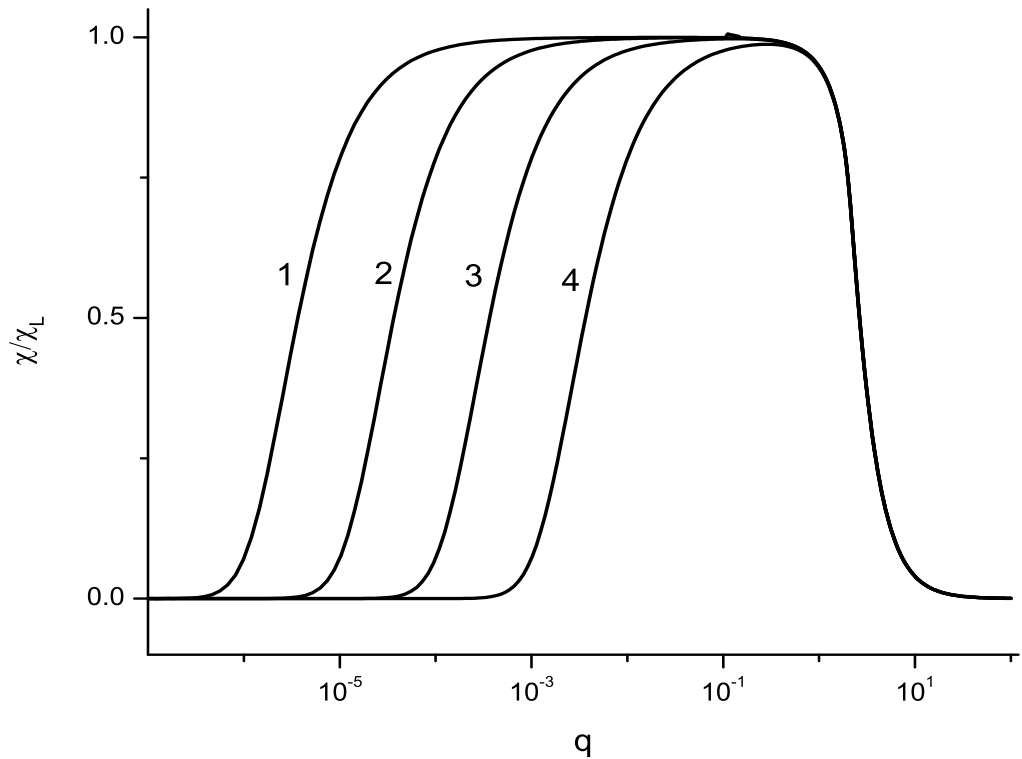}%[width=8.0cm, height=4.5cm]
\caption{Diamagnetic susceptibility
for the case $x=0\,(\omega=0)$ (a static limit), curves
$1,2,3,4$ correspond to values of parametre $y=10^{-6}, 10^{-5}, 10^{-4},
10^{-3}$ .
}\label{rateIII}
\end{flushleft}
\end{figure}

In this case expression for the diamagnetic
susceptibility is given by the formula (3.5). On fig. 1 there are
presented dependence plot of the dimensionless diamagnetic
susceptibility
(divided by Landau diamagnetic susceptibility)
from the dimensionless wave
numbers at various electron collisions frequencies.
At $k\sim k_F\,(q\sim 1)$  the diamagnetic
susceptibility
coincides with Landau diamagnetism. At decreasing of wave number
the reduction of a diamagnetic susceptibility is observed up to
disappearances at $k\to 0$.

To understand the mechanism of this phenomenon, we will consider
a semiclassical  picture of electron
movement in  a magnetic field.

At the finite values of quantity $k$ the greatest contribution
to the diamagnetic response bring  electrons with Larmor
radius  orbits $\sim 1/k $. Thus frequency of their rotation
on this orbit $\sim v_F k$. If collisional frequency
of electrons exceeds rotation frequency, i.e. $\nu>v_F k$,
than corresponding Bohr orbits collapses and restores
classical picture of electrons movement. But in classical plasma
diamagnetism is absent. It occurs  just at $q<y$.

So our result gives answer to the question \cite{Datta}:
"Whether the Landau diamagnetism
itself survives dissipation?" The answer is "no".
\begin{center}
  \bf 6. Conclusons
\end{center}

In the present work the kinetic description of Landau's
diamagnetism is given for degenerate collisionless and collisional
plasmas with use before the formula deduced  for electric conductivity
of quantum plasma. For collisionalless plasmas with the help
the kinetic approach the known formula of Landau diamagnetism
is deduced, also it is shown, that in collisional plasma diamagnetism
of degenerate electronic gas  is equal to zero.
Thereby the answer to a question on influence  of dissipation  on
Landau diamagnetism is given.
This question has been put in work \cite {Datta}.

\end{document}